\documentstyle[aps,preprint]{revtex}
\input psfig

\def\ltap{\raisebox{-.55ex}{\rlap{$\sim$}} \raisebox{.4ex}{$<$}}
\def\gtap{\raisebox{-.55ex}{\rlap{$\sim$}} \raisebox{.4ex}{$>$}}
\def\gsim{\mathrel{\gtap}}
\def\lsim{\mathrel{\ltap}}

\newcommand{\dpar}[2]{\frac{\partial #1}{\partial #2}}

\def\be{\begin{equation}}
\def\ee{\end{equation}}

\def\e{\mbox{e}}

\def\k{\mbox{\bf k}}
\def\n{\mbox{\bf n}}
\def\v{\mbox{\bf v}} 
\def\r{\mbox{\bf r}} 
\def\p{\mbox{\bf p}}
\begin{document}
\draft
\preprint{PURD-TH-97-02, OSU-TA-01/97, hep-ph/9701423}
\date{January 1997}
\title{Relic gravitational waves produced after preheating}
\author{S. Yu. Khlebnikov$^1$ and I. I. Tkachev$^{2,3}$}
\address{
${}^1$Department of Physics, Purdue University, West Lafayette, IN 
47907 
\\
${}^2$Department of Physics, The Ohio State University, Columbus, 
OH 43210\\ 
${}^3$Institute for Nuclear Research of the Academy of Sciences of 
Russia\\ 
Moscow 117312, Russia}

\maketitle

\begin{abstract}
We show that gravitational radiation is produced quite efficiently 
in interactions of classical waves created by resonant decay of a 
coherently oscillating field. For simple models of chaotic inflation
in which the inflaton interacts with another scalar field, we find 
that today's ratio of energy density in gravitational waves per octave 
to the critical density of the universe can be as large as
$ 10^{-12}$ at the maximal wavelength of order $10^{5}$  cm. 
In the pure $\lambda\phi^4$ model, the maximal today's wavelength 
of gravitational waves produced by this mechanism is of order 
$10^6$ cm, close to the upper bound of operational LIGO and TIGA 
frequencies. The energy density of waves in this model, though, is 
likely to be well below the sensitivity of LIGO or TIGA at such 
frequencies. We discuss possibility that in other inflationary models 
interaction of classical waves can lead to an even stronger gravitational 
radiation background.
\end{abstract}


\narrowtext
\section{Introduction}
Recent research in inflationary cosmology has attracted attention to 
highly non-equilibrium states created in a decay of a 
coherently oscillating field after the end of inflation.
These states could support a number of non-equilibrium 
phenomena,  such as non-thermal symmetry restoration \cite{effects}
and baryogenesis \cite{effects,bau}
shortly after or during the decay of the oscillating field.

In this paper we want to show that non-equilibrium
states produced by the decay of coherent oscillations of a field are
a quite efficient source of a stochastic background  of 
gravitational waves. 

There are several possible processes in the early universe capable to 
produce a stochastic background of relic gravitational waves. One is 
the parametric amplification of vacuum graviton fluctuations during
inflation  \cite{LG75}. This process is efficient on all frequency 
scales. 
Waves with lowest frequencies cause inhomogeneities of cosmic 
microwave
background \cite{RSV}. In conventional scenarios, this restricts the 
amplitude 
of high-frequency gravitational waves to be far below \cite{mt} the 
experimental 
limits accessible for direct detection experiments in near future ;
this conclusion changes in superstring motivated cosmologies
\cite{ss}. Another source of gravitational radiation is 
classical emission that accompanies collisions of
massive bodies. A natural source in this class in the early universe
is a strongly first-order phase transition when gravitational waves
are produced in collisions of bubbles of a new phase 
\cite{ptr1,TW,ptr2}, in particular the phase transition that
terminates first-order inflation \cite{TW}.
Gravitational radiation is also emitted during the decay of a cosmic 
string network \cite{cs}.

In this paper we discuss a new source of relic gravitational
waves. Decay of coherent oscillations of a scalar field 
can produce large, essentially classical fluctuations via
parametric amplification (resonance) \cite{GMM}. These classical 
fluctuations, which can be viewed either as classical waves 
traveling 
through the universe or, at least qualitatively, as quantum ``particles'' 
in states with large occupation numbers, interact with the oscillating 
background and each other. This interaction, which we call 
rescattering \cite{us},
is accompanied by gravitational radiation. That is the effect we want 
to estimate.

Favorable conditions for an effective parametric resonance in cosmology 
naturally appear in inflationary models \cite{KLS,resonance}.
We consider two types of simple inflationary 
models here. One type of models has two scalar fields with an interaction
potential of the form $g^2\phi^2 X^2/2$, 
and the resonance produces mostly fluctuations of a scalar field $X$
other than the field $\phi$ that oscillates (although subsequent 
rescattering processes produce large fluctuations of the field $\phi$ 
as well). We consider a range of moderate values of the coupling $g^2$ (see
below); in this case fluctuations are not suppressed too strongly by 
non-linear effects (cf. Refs. \cite{wide,resc,PR}).
In simplest models of this type, $\phi$ is the inflaton itself. 
For these models, we find that typically $\sim 10^{-5}$ of the total 
energy of the universe go into gravitational waves, at the time of 
their production. 
The minimal today's frequency $f_{\min}$ of these waves is 
typically of order
$10^{5}$ Hz, and today's spectral density  at this frequency can be 
as large as $10^{-12}$ of the critical density.

Another type of model we considered was the pure $\lambda\phi^{4}$ 
model of chaotic inflation. In this model, the minimal today's frequency 
$f_{\min}$ is 
of order $10^{4}$ Hz, close to the upper bound of the 
operational LIGO and TIGA frequencies \cite{LIGO}:
10 Hz $\alt f_{\rm LIGO}\alt 10^4$ Hz. 
We do not yet have efficient means of extrapolating our numerical results 
for today's spectral intensity to the minimal frequency of this model,
but we do not expect it to be above $10^{-11}$ of the critical 
density. That would be well below the sensitivity of LIGO or TIGA at 
frequencies of order $10^{4}$ Hz.  

We believe, however,
that in the absence of a commonly accepted specific inflationary 
model, it is premature to rule out possibility of experimental 
detection, even already by LIGO or TIGA, 
of gravity waves produced by the mechanism we consider here.
At the end of this paper,
we discuss possibility of a stronger background of 
gravitational waves in models with more fields or more complicated
potentials.

\section{General features of post-inflationary dynamics}
In scenarios where inflation ends by resonant
decay of coherent field oscillations,
post-inflationary dynamics has two rapid stages. 
At the first of these, called preheating \cite{KLS}, fluctuations of 
Bose 
fields interacting with the oscillating field grow exponentially 
fast, as a 
result of parametric resonance, and achieve large occupation numbers.
At the second stage, called semiclassical thermalization \cite{us}, 
rescattering of produced fluctuations smears out the resonance peaks 
in power spectra and leads
to a slowly  evolving state, in which the power spectra are smooth
\cite{us,wide,resc}. The system begins to exhibit chaotic 
behavior  
characteristic of a classical non-linear system with many degrees of 
freedom. In the course of subsequent slow evolution, the power
spectra propagate to larger momenta; we expect that eventually this 
process will lead to a fully thermalized state (which does not admit 
a semiclassical description).

In many models, we find that during the stage of semiclassical 
thermalization, or chaotization,
fluctuations grow somewhat beyond their values at the end of 
the resonance stage. In such cases,
the most effective graviton production takes place at the end of and
shortly after the chaotization stage. 

In order not to confine ourselves to any particular type of 
inflationary scenario, we will not assume that the oscillating field 
$\phi$ is the inflaton itself. 
(It is distinct from the inflaton, 
for example, in hybrid inflationary models \cite{hyb}.)
Let us denote the amplitude of the oscillating field $\phi$ at the end
of the chaotization stage as $\phi_{ch}$, the frequency of its 
oscillations 
as $m$, and the Hubble parameter at that time as $H_{ch}$. These same 
parameters at the end of inflation, when the oscillations start, will 
be denoted as $\phi(0)$, $m(0)$, and $H(0)$. 
In simplest models of chaotic inflation,
in which $\phi$ is the inflaton, $\phi(0)\sim M_{\rm Pl}$, and
$H(0)\sim m(0)$. Although we will use such models for 
illustrative purposes, our general formulas do not assume these 
conditions. Oscillations 
of $\phi$ cannot start unless $m(0)\gsim H(0)$, but on the other hand,
they can start at
$m(0)\gg H(0)$, if they have to be triggered by some other field.
Even if $H(0)\sim m(0)$, as in simple models of chaotic inflation, we 
still have
\begin{equation}
H_{ch}< H_{r}\ll m
\label{con}
\end{equation}
where $H_{r}$ is the Hubble parameter at the end of the resonance 
stage. This is because the frequency of oscillations 
redshifts slower (if at all) than the Hubble parameter. We will use
the condition (\ref{con}) in what follows.

We will consider models, in which oscillating field
$\phi$ interacts with a massless scalar field  $X$. The Lagrangians 
for these models are of the form
\begin{equation}
L={1\over 2} g^{\mu\nu}\partial_{\mu}\phi \partial_{\nu}\phi
+ {1\over 2}
g^{\mu\nu}\partial_{\mu} X \partial_{\nu} X - V(\phi,X)
\label{L}
\end{equation}
where 
\begin{equation}
V(\phi,X)=V_{\phi}(\phi)  + {1\over 2} g^2 \phi^2 X^2
\label{pot}
\end{equation}
and $V_{\phi}$ is a potential for the field $\phi$. We will consider 
two
types of $V_{\phi}$: $V_{\phi}={1\over 2} m^2\phi^2$ (massive $\phi$) 
and $V_{\phi}={1\over 4}\lambda\phi^4$ (massless $\phi$).
The effect we discuss is not limited to these particular models and 
exists 
in a wide variety of inflationary models that have a preheating stage.
Note that in the model with massless $\phi$,
the frequency $m$ of oscillations of $\phi_{0}$ at the end of 
chaotization is $m\sim \sqrt{\lambda} \phi_{ch}$.

The oscillating $\phi$ amplifies fluctuations 
of $X$ via parametric resonance. 
An important parameter in the problem is the resonance 
parameter
$q$. Depending on the choice of the potential for $\phi$ (see above),
$q=g^2\phi^2(0)/4m^2$ (massive $\phi$) or $q=g^2/4\lambda$
(massless $\phi$).

In the case of massive $\phi$, it is useful to introduce, 
in addition to the resonance
parameter $q$, the redshifted resonance parameter at the end of the 
resonance stage 
\begin{equation}
q_{r}=q\frac{{\bar \phi}^{2}(t_{r})}{\phi^{2}(0)}
\label{qeff}
\end{equation}
where $t_{r}$ is the time corresponding to the end of the resonance 
stage, and ${\bar \phi}(t_{r})$ is the amplitude of oscillations of 
$\phi_{0}$ at that time. In the model with massive $\phi$ and massless
$X$, parametric resonance can fully develop in an expanding universe if 
$q_{r}\gsim 1$ \cite{wide}. Similarly, we can introduce
\begin{equation}
q_{ch}=\frac{q \phi_{ch}^{2}}{\phi^{2}(0)}
\label{qch}
\end{equation}
For uniformity of notation, 
we will sometimes use $q_{r}$ or $q_{ch}$ instead of $q$ in the case of 
massless 
$\phi$; there is no difference between the three of these in that case.

Resonant production
is most effective for fluctuations of fields that couple to $\phi$ 
not 
too weakly but also not too strongly, those with $q_{r}\sim 1$.
For $q_{r}\gg 1$, the maximal size of $X$ fluctuations is significantly
suppressed by non-linear effects
\cite{wide,resc,PR}. Because, for instance, superstring
models predict a plethora of scalar fields, we expect that some of 
those 
will have couplings in the optimal range. So, in what follows we 
consider moderate values of $q_{r}$, $1\lsim q_{r} \leq 100$.

In the model with massive $\phi$,
the oscillating zero-momentum mode $\phi_{0}$
drops rapidly at the end of the chaotization stage, and 
all  its energy at that time is transferred to fluctuations \cite{resc}. 
The variance of $\phi$, $\langle (\delta\phi)^{2} \rangle$ at the end
of chaotization is thus much larger than $\phi_{ch}^{2}$. This does 
not happen in the model with massless $\phi$. 
So, in what follows we consider 
two cases: $\langle (\delta\phi)^{2} \rangle \lsim \phi_{ch}^{2}$ and
$\langle (\delta\phi)^{2} \rangle \gg\phi_{ch}^{2}$.

For $q\gsim 1$ in the model where a massless $\phi$ interacts with $X$, 
parametric resonance for $\phi$ itself is  
insignificant. To study a case different in that respect,
we consider also the pure $\lambda\phi^{4}/4$ model, in which
$\phi$ decays solely due to self-coupling ($g^{2}=0$ in Eq. (\ref{pot})).

\section{Calculational procedures}
Because the time scale of processes that give rise to gravitational 
radiation is much smaller than the time scale of the expansion of the 
universe, $H_{ch}^{-1}$, the energy of gravitational waves can be 
approximately computed starting from the well-known formulas for flat 
space-time. Total energy of gravitational waves radiated in direction 
$\n=\k/\omega$ 
in flat space-time is \cite{Weinberg}
\be
\frac{dE}{d\Omega} = 2 G \Lambda_{ij,lm}(\n)
\int_0^{\infty} \omega^2 T^{ij*}(\k,\omega) T^{lm}(\k,\omega) d\omega
\label{for}
\ee
where $T^{ij}(\k,\omega)$ are Fourier components of the stress
tensor, and $ \Lambda_{ij,lm}$ is a projection tensor made of the 
components of $\n$ and Kronecker's deltas.

For models with two fields ($\phi$ and $X$) with potentials 
(\ref{pot}) and moderate $q_{r}$,
we will present both analytical estimates and numerical calculations
based on Eq. (\ref{for}). For analytical estimates, we choose a 
specific process---gravitational bremsstrahlung that accompanies 
creation and annihilation of fluctuations of the field $\phi$. This is 
not the only process that can produce a significant amount of 
gravitational radiation.
For example, because the mass of $X$ fluctuations oscillates due to 
interaction with $\phi_{0}$, the collection of $X$ fluctuations works 
(until $\phi_{0}$ decays completely) as a gravitational antenna. 
The reason why we concentrate on bremsstrahlung from $\phi$ is that
it is a significant source of 
gravity waves with {\em small frequencies}, which have the best 
chance 
to be detected today.
Indeed, we will see that both the frequency dependence and the 
overall 
magnitude of the effect are reasonably close to those of the {\em 
full} 
intensity for small frequencies that we obtain numerically. So, 
bremsstrahlung from $\phi$  appears to be at least one of the main 
sources of gravitational radiation with small frequencies in the 
states we consider here.

Numerical calculations were done for simple models of chaotic 
inflation, in which $\phi$ is the inflaton itself. The calculations
were done in conformal time, in which the evolution of the system is 
Hamiltonian (for more detail, see Ref. \cite{wide}), and then
rescaled back to the physical time. We computed directly Fourier 
transforms
$T^{ij}(\k,\omega)$ over successive intervals of conformal time, the 
duration of which was taken small compared to the time
scale of the expansion yet large enough to accommodate the minimal 
frequency of gravity waves we had in these calculations. Intensities 
from different intervals were summed up with the weight that
takes into account expansion of the universe (see more on that below).
Being a finite size effect,
the minimal frequency of our numerical calculations was much larger 
than the actual minimal frequency of gravity waves, $\omega_{\min}\sim 
H_{ch}$. In the models with two fields ($\phi$ 
and $X$), the spectrum obtained numerically 
is approximated reasonably well by the bremsstrahlung spectrum at 
small frequencies. This allows us to extrapolate the numerical results 
to $\omega\sim \omega_{\min}$ using the frequency dependence of 
bremsstrahlung.

For the pure $\lambda\phi^{4}/4$ model, the analytical method that we use 
to estimate the intensity of radiation does not apply at frequencies for 
which we have numerical data. So, we do not 
have efficient means of extrapolating our numerical results 
to $\omega_{\min}$. For this model, we 
contented  ourselves with numerical simulations.

\section{Analytical estimates}
In general, $X$ and $\phi$ fluctuations produced by parametric 
resonance scatter off the 
homogeneous oscillating background (condensate) of $\phi$,
knocking $\phi$ out of the condensate and into modes with non-zero 
momenta. In cases when 
resonance amplifies mostly fluctuations of $X$,
the scattering process can be viewed as a two-body decay of $X$,
$X_{\k}\to X_{\k'}+\phi_{\p}$, in the time-dependent background field
of the condensate. It can also be thought of as evaporation
of the inflaton condensate. There is also the inverse process 
(condensation).

As we noted in the introduction, one can describe the states we are 
considering either as collections of interacting classical waves, or
as collections of ``particles'' in modes with large occupation 
numbers. For estimating the intensity of bremsstrahlung that accompanies
creation and annihilation of fluctuations of the field $\phi$, the 
description of these fluctuations in terms of particles is more convenient. 
For the notion of 
particle, i.e an entity moving freely between collisions, to have
anything but purely qualitative meaning, the energy of a freely moving
``particle'' should not be modulated too strongly by its interactions with 
the background.

In the models (\ref{L}) with $q_{r}\gsim 1$, fluctuations of $\phi$ are 
reasonably well described as particles, even in the case of massless
$\phi$ where they are coupled to the oscillating $\phi_{0}$. Indeed,
typical momenta $p$ of $\phi$ fluctuations at the 
end of the chaotization stage are of order $m$, the frequency of 
oscillations of $\phi_{0}$ at that time. The frequency of 
oscillations of a mode with momentum $p\sim m$ (the would-be energy of a 
particle) is only moderately modulated by its coupling to the
oscillating $\phi_{0}$. 
Note also that for $q_{r}\gsim 1$, 
the Hartree correction to the frequency squared of $\phi$, $g^2 X^2$,
does not exceed $m^{2}$ itself \cite{resc}, so interaction with $X$ 
also does not modulate frequencies of fluctuations of $\phi$ too much.

So, let us consider fluctuations of $\phi$ as particles, neglecting 
modulations of their energy, and estimate the intensity 
of gravitational bremsstrahlung emitted by these particles in the 
scattering processes described above. We will assume that 
bremsstrahlung
from $\phi$ is at least one of the main sources of gravity waves with 
small frequencies (we will specify the notion of ``small'' 
frequency below). For estimation purposes, we will neglect its 
interference 
with possible small-frequency radiation from other sources. 

To describe radiation with small enough frequencies, we can
regard the particles---quanta of $\phi$---as semiclassical
wave packets. This can be seen by considering
the Fourier transform of the stress tensor of a single classical 
particle. 
Using identity transformations, we write that stress tensor as
\begin{equation}
\omega T_{1}^{ij}(\k,\omega)=\frac{1}{2\pi i}
\int_{-\infty}^{\infty}  \frac{p^i p^j}{p^0(1-\n\v)} \dpar{}{t}
\e^{i\omega t-i{\bf kr}} dt
\label{stress}
\ee
where $p^{\mu}$, $\r$, and $\v$ are the particle's time-dependent
four-momentum, position, and velocity. Uncertainty that we can tolerate
in the position of a particle, to be able to use this formula,
is of order $1/k$. Uncertainty we can tolerate in momentum 
is of order of the momentum spread among the particles, $\Delta p$.
Thus, the semiclassical approximation applies when 
$\omega=k\ll \Delta p$. In the models with two fields, at the end of 
chaotization, $\Delta p \sim m$, so Eq. (\ref{stress}) applies for 
$\omega\ll m$.  

On the other hand, in the pure $\lambda\phi^{4}/4$ model, fluctuations 
do not grow during chaotization, so for the most efficient 
graviton production we should consider the end of the resonance stage. 
The power spectrum of fluctuations at the end of resonance, and long 
into chaotization, is concentrated in rather narrow peaks, so that 
$\Delta p \ll m$. 
Then, Eq. (\ref{stress}) applies only at quite low frequencies, much 
lower than those we had in numerical simulations.
As a result, for this model, we could not check our 
analytical estimates against numerical simulations.

In the rest of this section we consider the models with two fields.
For $\omega\ll m$, not only we can use the semiclassical expression 
for the stress tensor of $\phi$, but in addition
each act of evaporation (or condensation) of a $\phi$ particle
can be regarded as instantaneous.
So, we can use the small frequency approximation,
familiar from similar problems in electrodynamics \cite{LL}. 
It amounts to integrating by parts in (\ref{stress}) and then replacing 
the exponential with unity.
After that, the integral is trivially taken and depends
only on the particle's final (or initial) momentum.

To obtain the power ${\cal P}$
radiated by a unit volume, we substitute the small-frequency
limit of Eq. (\ref{stress}) into Eq. (\ref{for}), multiply the result by 
the rate at which $\phi$ fluctuations are created (or destroyed), 
and integrate with the occupation numbers $n_{\phi}(p)$
 of $\phi$ fluctuations.
The rates of the creation and annihilation processes are almost
equal (see below). With both processes taken into account, we obtain
\begin{equation}
\frac{d{\cal P}}{d\omega}  \approx 4 G \int_{0}^{\infty} n_{\phi}(p)
{\cal R}(p) F(v) \frac{p^{4}dp}{(2\pi)^{3}}
\label{power}
\end{equation}
where $P_{\phi}(p)$ is the power spectrum of $\phi$, ${\cal R}(p)$
is the evaporation (condensation) rate, $v=p/p^{0}$, and the function
\begin{equation}
F(v)=\frac{4}{v^{2}} \left( 2- \frac{4}{3} v^{2} -
\frac{1-v^{2}}{v}\ln\frac{1+v}{1-v} \right)
\label{F}
\end{equation}
arises after integration over direction of particle's momentum;
we have assumed that both the power spectrum and the rate depend 
only on the absolute value of momentum.
Notice that the spectrum (\ref{power}) 
is $\omega$ independent, as characteristic of small-frequency 
bremsstrahlung.

To estimate the rate ${\cal R}$, we will consider, qualitatively,
$X$ fluctuations also as ``particles'' characterized by their own
occupation numbers $n_{X}(k)$. This is not a strictly defensible 
view, as $X$ fluctuations are strongly coupled to the oscillating 
background, but it should do for estimation purposes. 

Consider first
the case $\langle (\delta\phi)^{2} \rangle \lsim \phi_{ch}^{2}$.
The rate ${\cal R}(p)$ can then be estimated as
\begin{equation}
{\cal R}(p) \sim \frac{g^{4} \phi_{ch}^{2}}{m}                   
\int \frac{d^{3}k}{(2\pi)^{3}}
\frac{n_{X}(\mbox{\bf k}) n_{X}(\mbox{\bf k}-\mbox{\bf p})}
{8\omega_{\mbox{\bf k}} \omega_{\mbox{\bf k}- \mbox{\bf p}} p^{0}}
\label{rate}
\end{equation}
where a factor of $1/m$ appears instead of the usual energy delta 
function because energy of the ``particles'' participating in the
process is not conserved, due to time-dependence of $\phi_{0}$.
This factor estimates the time scale at which energy non-conservation 
sets in; in our case, it is the period of the oscillations of 
$\phi_{0}$.

Notice that the rate (\ref{rate}) is larger than the net kinetic rate,
which would enter a kinetic equation for $\phi$, by a factor of 
order of a typical occupation number of $X$. 
The net kinetic rate is the difference between two 
(large) numbers---the rate at which collisions supply particles to
a given mode and the rate at which they remove them. But each 
collision
is accompanied by bremsstrahlung, so ${\cal R}$ is not 
the net kinetic rate but a rate of the ``in'' and ``out'' 
processes separately.

For moderate $q_{r}$, we can use the following estimates (cf. Refs.
\cite{us,wide,resc}) : 
$p\sim k \sim \omega_{k} \sim m$,
$n_{\phi}\sim n_{X}\sim 1/g^{2}$. Using these estimates, we obtain
${\cal R}(p) \sim \phi_{ch}^{2}/m$, for $p\sim m$, and
\begin{equation}
\frac{d{\cal P}}{d\omega} \sim 
\frac{m^{4}\phi_{ch}^{2}}{g^{2}M_{\rm Pl}^{2}}
\sim \frac{m^{2}\phi_{ch}^{4}}{M_{\rm Pl}^{2}}
\label{power1}
\end{equation}
where in the last relation we have used 
$g^{2}\sim q_{ch}m^{2}/\phi_{ch}^{2}$ and dropped a factor of 
$1/q_{ch}$.

An estimate for the total energy density of 
gravitational waves is obtained by multiplying the power
${\cal P}$ by the time $\Delta t$ during which the radiation was 
substantial. For the ratio of $\rho_{\rm GW}$ per octave to the 
total energy density of the universe after chaotization, we obtain
\begin{equation}
\left(
\frac{1}{\rho_{\rm tot}} \frac{d \rho_{\rm GW}}{d\ln\omega} 
\right)_{ch}
\sim
\frac{m^{2} \phi_{ch}^4}{M_{\rm Pl}^2 \rho_{\rm tot}} 
\omega \Delta t
\sim \frac{\phi_{ch}^2}{M_{\rm Pl}^2} \omega \Delta t
\label{enden}
\end{equation}
where in the last relation
we used $ \rho_{\rm tot} \sim m^2\phi^2_{ch}$ .
The time during which the radiation 
was substantial is determined by the redshift, $\Delta t \sim 
H_{ch}^{-1}$.

For the opposite case
$\langle (\delta\phi)^{2} \rangle \gg \phi_{ch}^{2}$, we need
to replace $\phi_{ch}^{2}$ in the estimate (\ref{rate}) by  
$\langle (\delta\phi)^{2}\rangle$. Now using 
$\langle (\delta\phi)^{2}\rangle \sim m^{2}/g^{2}$ and
$\rho_{\rm tot}\sim m^{2} \langle (\delta\phi)^{2}\rangle$, we see 
that we need to make the same replacement in Eq.(\ref{enden}).

These estimates apply at the end of the chaotization
stage and extend into the infrared up to frequencies 
of the order of the horizon scale at that time, $\omega\sim H_{ch}$.
(We will present estimates for today's $\rho_{\rm GW}$ below.)

As an example, consider the model with massive $\phi$ in which
$\phi$ is the inflaton itself, and $m=10^{-6}M_{\rm Pl}$.
Take $q_{r}\sim 1$, which in this case corresponds 
to $q\sim 10^{4}$ \cite{wide}. The inflaton completely decays into
fluctuations during the chaotization stage, and at the end of it
$\langle (\delta\phi)^{2}\rangle\sim 10^{-6}M_{\rm Pl}$ \cite{resc}. 
Using $\langle (\delta\phi)^{2}\rangle$ in place of $\phi_{ch}^{2}$
in Eq. (\ref{enden}), we obtain
$\rho_{\rm tot}^{-1}d \rho_{\rm GW}/d \ln\omega\sim 10^{-6}$
at $\omega\sim H_{ch}$. 

Our analytical estimates are rather crude, as we were somewhat 
cavalier 
with numerical factors. They do bring out, however, not only the 
frequency dependence of the effect but to some extent also 
its overall magnitude. Notice, for example, that the last estimate in
Eq. (\ref{enden}) does not contain powers of $g^{2}$ or $m^{2}$ (these 
two can be expressed through each other and $\phi_{r}^{2}$,
using the condition $q_{r}\sim 1$). These quantities, if 
present, would change the estimate by many orders of magnitude,
in drastic disagreement with our numerical results.

\section{Background gravitational radiation today}
Let us now translate the above estimates into estimates for 
gravitational  background radiation today.
It is important  that gravitational
radiation produced after preheating has not interacted with 
matter since then \cite{ZN}.

First, let us estimate the physical wavelength today,  
$l= 2\pi/k_0$, that corresponds to a wave vector $k_{ch}$ at
the end of the chaotization stage.
We have $k_0=k_{ch} a_{ch}/a_0$, where $a_0$ is the scale factor today.
Thermal equilibrium was
established at some temperature $T_*$, and the universe was radiation
dominated at that time. This happened when
relevant reaction rates became equal to the expansion rate.
We can find the Hubble parameter at that time as $H_*^2=8\pi G 
\rho_*/3$, 
where $\rho_*=g_{*}\pi^2T_*^4/30$, and $g_*\equiv g(T_*)$;
$g(T)$  stands for the effective
number of ultrarelativistic degrees of freedom at temperature $T$.
The expansion factor from $T=T_*$ down to $T=T_0 \approx 3 K$
is given by $a_*/a_0 = (g_0/g_*)^{1/3} (T_0/T_*)
=(g_0^{1/3}/g_*^{1/12})(8\pi^3/90)^{1/4}(T_0/\sqrt{H_*M_{\rm Pl}})$,
where $g_{0}\equiv g(T_{0})$. Depending
upon model parameters, the universe could expand from $T=T_{ch}$
to $T=T_*$ as matter or radiation dominated. We can write
$H_*=H_{ch}(a_{ch}/a_*)^\alpha$, where $\alpha=2$ for the radiation 
dominated
case and $\alpha =3/2$ for a matter dominated universe.
We obtain $k_0=k_{ch} a_{ch}/a_0=k_{ch} (a_{ch}/a_*)(a_*/a_0)
\approx 1.2 k_{ch} (H_{ch} M_{\rm Pl})^{-1/2} (a_{ch}/a_*)^{1-\alpha/2} T_0$.
The model dependent factor $(a_{ch}/a_*)$ does not enter this relation
in the radiation dominated case and enters in power $1/4$ 
for a matter dominated universe, i.e. this factor gives a not very 
important correction (at most an order of magnitude).
For today's wavelength, corresponding to wave vector $k_{ch}$ at the end
of chaotization, we find
$l = 2\pi/k_0 \approx 0.5\, (M_{\rm Pl} H_{ch})^{1/2} k_{ch}^{-1} 
(a_*/a_{ch})^{1-\alpha/2}$ cm. 

The smallest wave vector of radiation that could be 
produced at the end of chaotization is of order $H_{ch}$. 
The corresponding maximal today's wavelength $l_{\max}$ will fall 
into the range of the
LIGO detector, i.e. $\l_{\max}>3\times 10^{6}$ cm, when
$H_{ch}< 10^{5}$--$10^{6}$ GeV. 

For illustration, let us consider
simple models of chaotic inflation with potentials (\ref{pot}), 
where the oscillating field $\phi$ is the inflaton itself.
The Hubble parameter at the end of the chaotization stage in these 
models can be extracted from numerical integrations of 
Refs. \cite{us,wide,resc}. For example, for massless inflaton 
we get $H_{ch}/M_{\rm Pl} = (3\lambda/ 2\pi)^{1/2} \tau_{ch}^{-2}$, where
$\tau_{ch}$ is conformal time at the end of the chaotization stage.
For $\lambda=10^{-13}$ and, say, $q=30$ we get (see Fig. 1) 
$\tau_{ch} \approx 125$. This gives 
$H_{ch}/M_{\rm Pl}\approx 1.4\times 10^{-11}$ and 
$l_{\max}\approx 1.3\times 10^5$ cm. Another example
is the case of massive inflaton with $m = 10^{-6} M_{\rm Pl}$ 
and $q=10^4$. Here we find
$H_{ch}/M_{\rm Pl}\approx 2 \times 10^{-9}$.
This gives $l_{\max}$ in the range $10^4$--$10^{5}$ cm. 

An interesting case is that of the pure $\lambda\phi^{4}/4$ model
($g^{2}=0$ in (\ref{pot})). In this case, fluctuations do not grow 
during chaotization (they actually decrease due to redshift), so the 
most 
efficient production of gravity waves takes place after the end of 
the 
resonance (preheating) stage. So, instead of $H_{ch}$ in the above 
formulas we use $H_{r}$, 
the Hubble parameter at the end of resonance. We have 
$H_{r}/M_{\rm Pl} = (3\lambda/ 2\pi)^{1/2} \tau_{r}^{-2}$, where 
$\tau_{r}$ is conformal time at the end of resonance. In this model,
resonance develops slower that in models with $q_{r}\gsim 1$.
As a result, $H_{r}$ is smaller, and today's maximal wavelength is 
larger. For $\lambda=10^{-13}$, we get $\tau_{r} \approx 500$ 
\cite{us}, 
and $H_{r}/M_{\rm Pl} \approx 9\times 10^{-13}$,
which gives, for radiation produced at that time, 
$l_{\max}\approx 5 \times 10^5 $ cm. Note, that by the time
$\tau \approx 1000$ the inflaton still has not decayed in this model, 
fluctuations are still highly non-equilibrium and gravity waves 
generation
will continue even at later time. This will increase $l_{\max}$ by at 
least another factor of two bringing it close 
to the upper boundary of operational LIGO and TIGA frequencies.

Now, let us estimate today's intensity of radiation. The today's 
ratio of energy in gravity waves to that in radiation is
related to $(\rho_{\rm GW}/\rho_{\rm tot})_{ch}$ via
\begin{equation}
\left( \frac{\rho_{\rm GW}}{\rho_{\rm rad}} \right)_{0}=
\left(\frac{\rho_{\rm GW}}{\rho_{\rm tot}} \right)_{ch}
\left( \frac{a_{ch}}{a_{*}} \right)^{4-2\alpha}
\left( \frac{g_{0}}{g_{*}} \right)^{1/3}
\label{rel}
\end{equation}
For the often used parameter 
$\Omega_{g}(\omega) \equiv (\rho_{\rm crit}^{-1}d\rho_{\rm
GW}/d\ln\omega)_{0}$, where $\rho_{\rm crit}$ is the critical energy 
density, we obtain, using (\ref{enden}),
\begin{equation}
\Omega_g (\omega) h^2 \sim
\Omega_{\rm rad} h^2 \frac{\phi_{ch}^{2}}{M_{\rm Pl}^{2}}
\frac{\omega}{H_{ch}}
\left( \frac{a_{ch}}{a_{*}} \right)^{4-2\alpha}
\left( \frac{g_{0}}{g_{*}} \right)^{1/3}
\label{omest}
\end{equation}
where $\Omega_{\rm rad}$ is today's value of the ratio
$\rho_{\rm rad}/\rho_{\rm crit}$: 
$\Omega_{\rm rad} h^2 = 4.31 \times 10^{-5}$ \cite{KT}.
When the oscillating zero-momentum mode decays completely
during the stage of chaotization, $\phi_{ch}^{2}$ in (\ref{omest})
is replaced by $\langle(\delta\phi)^{2}\rangle_{ch}$. 

For example, in the case of massless inflaton with $\lambda=10^{-13}$ 
and $q=30$, we have $\phi_{ch}^2\sim 10^{-5} M_{\rm Pl}^2$, and with 
$g_*/g_0 \sim 100$ the estimate
(\ref{omest}) gives $\Omega_g h^2 \sim 10^{-10}$ at 
$l_{\rm max}\sim 10^5$ cm.

\section{Numerical results}
We have studied  numerically gravitational radiation in the model
where $\phi$ is the massless inflaton with 
$V_{\phi}=\lambda\phi^{4}/4$,
$\lambda=10^{-13}$ , interacting with a massless field $X$ according 
to 
Eq. (\ref{pot}). We have used the fully non-linear method of Ref. 
\cite{us}.
We have solved equations of motion on a lattice $128^3$ in a box with
periodic boundary conditions.

The model is classically conformally invariant, so by going to 
conformal variables, the equations of motion 
can be reduced to those in flat space-time. 

Time evolution of the variances $\langle X^{2} \rangle$ and
$\langle (\delta\phi)^2 \rangle$ (in rescaled untis) for 
resonance parameter $q=30$ is shown in Fig. 1. The rescaled
conformal time $\tau$ is related to time $t$ by
$\sqrt{\lambda}\phi(0) dt= a(\tau)d\tau/a(0)$. 
The rescaled conformal fields $\chi$ and $\varphi$ are related to the 
original fields by 
$X=\chi\phi(0) a(0)/a(\tau)$ and $\varphi =\phi \phi(0) a(0)/a(\tau)$.
In this model, $\phi(0)\approx 0.35 M_{\rm Pl}$ and 
$a(\tau)/a(0)\approx 0.51\tau+1$ \cite{wide}.

We see that for the above values of parameters, 
parametric resonance ends at $\tau\approx 73$.
The resonance stage is followed by a plateau 
(cf. Refs. \cite{wide,resc}). At the plateau, the variances
of fluctuations do not grow, but an important restructuring of the 
power spectrum of $X$
takes place. The power spectrum of $X$ changes from being dominated 
by a resonance peak at some non-zero momentum to being dominated 
by a peak near zero. 
(For some $q$, though, the strongest peak is close to 
zero already during the resonance.)
When the peak near zero becomes strong enough, the growth of 
variances resumes (in Fig. 1, that happens at $\tau\approx 84$), and 
the system enters the chaotization stage.

The power spectrum of 
$\phi$ during the time interval of Fig. 1 is shown in Fig. 2. The 
rescaled comoving momentum $k$ is related to physical momentum
$k_{\rm phys}$ as $k_{\rm phys}=\sqrt{\lambda}\phi(0) a(0) k/a(\tau)$.
Note that in the range  $1\lsim k \lsim 10$, the power spectrum
is approximately a power law, as characteristic of Kolmogorov spectra
\cite{Zakharov}.

Interaction of the fields with gravity is not conformal, and the 
flat-space-time formula (\ref{for}) for the energy of gravity waves
can be used only after we make 
an approximation. The approximation replaces the actual expanding 
universe with a sequence of static universes. Specifically, conformal 
time was divided into steps of $\Delta \tau=2\pi L$, where $L$ is the 
size
of the integration box, and at each step the 
physical variables (fields, frequency, and momenta) 
were obtained from the conformal ones, using for
$a(\tau)$ the actual scale factor taken at the middle point of the
step. The energy of gravitational waves was computed at each step using
(\ref{for}), with the corresponding physical variables. Then, 
energies from all the steps were summed up.

Today's $\Omega_{g}$, in physical units, that had been
accumulated by conformal time $\tau=200$ is shown in Fig. 3 
for two values of $q$: $q=30$ (dashed line) and $q=105$ (dotted 
line). 
The solid line corresponds to the pure $\lambda \phi^4$ model 
(no interaction with $X$ field) and includes contributions from times 
up to
$\tau =1600$. In the latter case, the peaks in $\Omega_g$ at frequencies
$f \agt 4 \times 10^7$ Hz, seen in the figure, are correlated with 
peaks in 
the power spectrum of $\phi$ at early stages of chaotization, cf. 
Ref. \cite{us}. Since we included contributions to $\Omega_g$ from 
times 
long after the time when the peaks in the power spectrum disappeared, 
and yet 
the peaks in $\Omega_g$ had not been washed out, we suggest that 
in this model the peaks
in $\Omega_g$ are a feature potentially observable at present.
Observation 
of such peaks could select a particular model of inflation.

We see also that in the model with two interacting fields ($\phi$ and 
$X$),
the linear $\omega$ dependence at small $\omega$, characteristic 
of bremsstrahlung, is overall well born out, both for $q=30$ and 
$q=105$. 
The minimal today's frequency in this model is $f_{\min}\sim 10^{5}$ 
Hz. Extrapolating the results of Fig. 3 to that minimal frequency using 
the linear law, we obtain the magnitude of $\Omega_{g}h^2$ 
of order $10^{-12}$ for $q=30$ and of order $10^{-13}$ for $q=105$. 
Recall that the analytical estimate (\ref{enden}) 
(using $\phi_{ch}^{2}\sim 10^{-5} M_{\rm Pl}^{2}$) gave 
$\Omega_{g}h^2$ of order $10^{-10}$ at the minimal frequency.

In the pure $\lambda\phi^4/4$ model, we could not discern any simple 
pattern
of frequency dependence for $\Omega_g$ at small frequencies. This is
consistent with our discussion of this model in Sect. IV: 
if the  linear
frequency dependence sets in at all in this model, that happens only 
at frequencies much 
smaller than those in Fig. 3. This makes it difficult 
for us to extrapolate our numerical results for this model to the maximal 
today's wavelength, $l_{\max}\sim 10^6$ cm, or, equivalently, minimal 
frequency $f_{\min}$ of order $10^4$ Hz. 
We have no reason to suppose, however, that $\Omega_g h^2$
can actually increase to small frequencies, beyond its 
value of order $10^{-11}$ at $f\sim 10^6$ Hz.

To confirm that Fig. 3 is not a numerical artifact, we have made
runs in which the interaction was switched off,
i.e. $g^{2}$ was set to zero, at $\tau=100$, for the case $q=30$. To 
exclude also the effect of self-interaction, the term $\lambda \phi^4$ was 
replaced at that time by $m^2 \phi^2$. The 
system then becomes a collection of free particles (plus the oscillating
zero-momentum background) and, if solved exactly in infinite space, should 
not radiate.
In our simulations, the intensity of radiation during the interval from
$\tau=100$ to $\tau=125$ was three orders of magnitude smaller than
it was during the same time interval with the interaction present.

\section{Conclusion}
We have shown that gravitational radiation is produced quite efficiently
in interactions of classical waves created by resonant decay of a
coherently oscillating field. For simple models of chaotic inflation
in which the inflaton interacts with another scalar field, we find
that today's ratio of energy density in gravitational waves per octave
to the critical density of the universe can be as large as
$ 10^{-12}$ at the maximal wavelength of order $10^{5}$  cm.
In the pure $\lambda\phi^4$ model, the maximal today's wavelength
of gravitational waves produced by this mechanism is of order
$10^6$ cm, close to the upper bound of operational LIGO and TIGA
frequencies. The energy density of waves in this model
likely to be well below the sensitivity of LIGO or TIGA at such
frequencies. 

In other types of inflationary models (or even with other parameters)
the effect can be much stronger. We do not exclude that among these there 
are cases in which it can be observable already by LIGO or TIGA.
The relevant situations are:
\begin{enumerate}
\item At some values of the coupling constant $g^2$ (or the resonance 
parameter $q$), the most resonant momenta are close to $k=0$. 
In the model with massless inflaton this happens, for example, for
$q=100$ (and does not happen for $q=30$ or $q=105$, which we 
discussed so far;
in these cases, the most resonant momenta are at $k\sim 1$).
The lowest frequency that we had in the box in our numerical 
simulations for $q=100$ was $f \approx 8\times 10^6$ Hz. At that
frequency, $\Omega_g$ for $q=100$ was almost two orders of magnitude
larger than for $q=105$. 
In addition, the entire spectrum of gravitational 
waves appears to be shifted to the left with respect to the spectra
shown in Fig. \ref{fig:Fig3}. 
Cases when resonance is ``tuned'' to be
close to $k=0$ deserve further study. Question remains, how large,
in such cases, the
intensity of gravitational waves can be at the horizon scale, $H_{ch}$.
\item It is important to consider in detail models of hybrid inflation 
\cite{hyb}, where the oscillating field need not be the inflaton 
itself, and so the frequency of the oscillations may be unrelated to the
inflaton mass.
\item In models where large fluctuations produced at preheating cause 
non-thermal phase transitions, as suggested in Ref. \cite{effects},
domains or strings can form.
A large amount of gravitational radiation can be produced in 
collisions of domain walls, in a way somewhat similar to how it happens 
\cite{TW,ptr2} in models of first-order inflation, or in decays of a 
string network, cf. Refs. \cite{cs}. In particular, in cases when 
domains are formed,
intensity of gravitational radiation at the horizon scale, at the moment 
when the domain structure disappears, is expected to be much larger 
than in cases without domains. 
\end{enumerate}

We thank L. Kofman and A. Linde for discussions 
and comments. The work of S.K. was supported
in part by the U.S. Department of Energy under Grant 
DE-FG02-91ER40681 
(Task B), by the National Science Foundation under Grant PHY 
95-01458, 
and by the Alfred P. Sloan Foundation. The work of I.T. was supported
by DOE Grant DE-AC02-76ER01545 at Ohio State.

\begin{figure}
\psfig{file=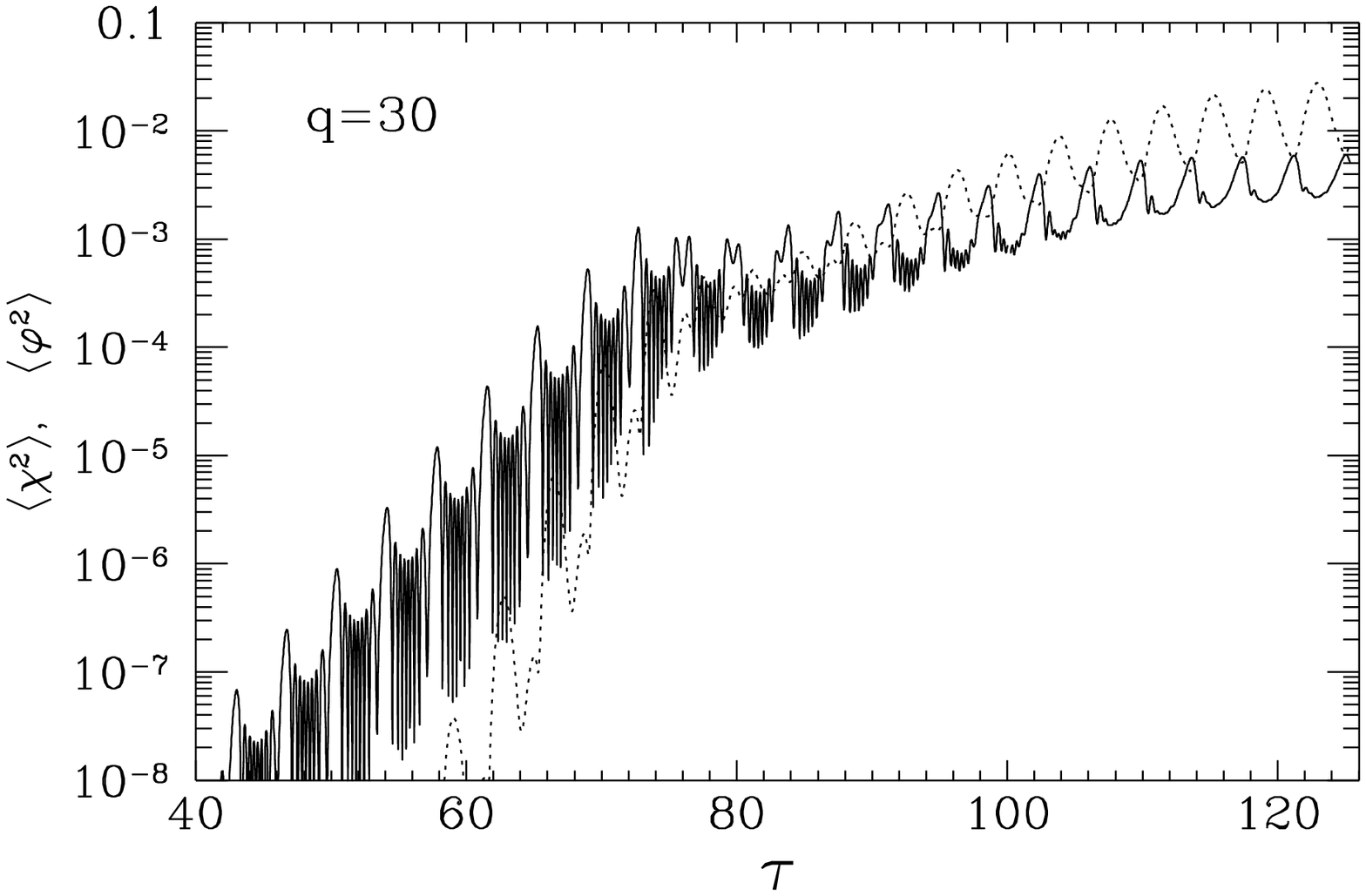,height=3.6in,width=5.6in}
\caption{Variances of fields $X$ (solid curve) and $\phi$ 
(dotted curve) as functions of conformal time in the model with massless 
inflaton for $\lambda=10^{-13}$ and $q=30$. }
\label{fig:Fig1}
\end{figure}

\begin{figure}
\psfig{file=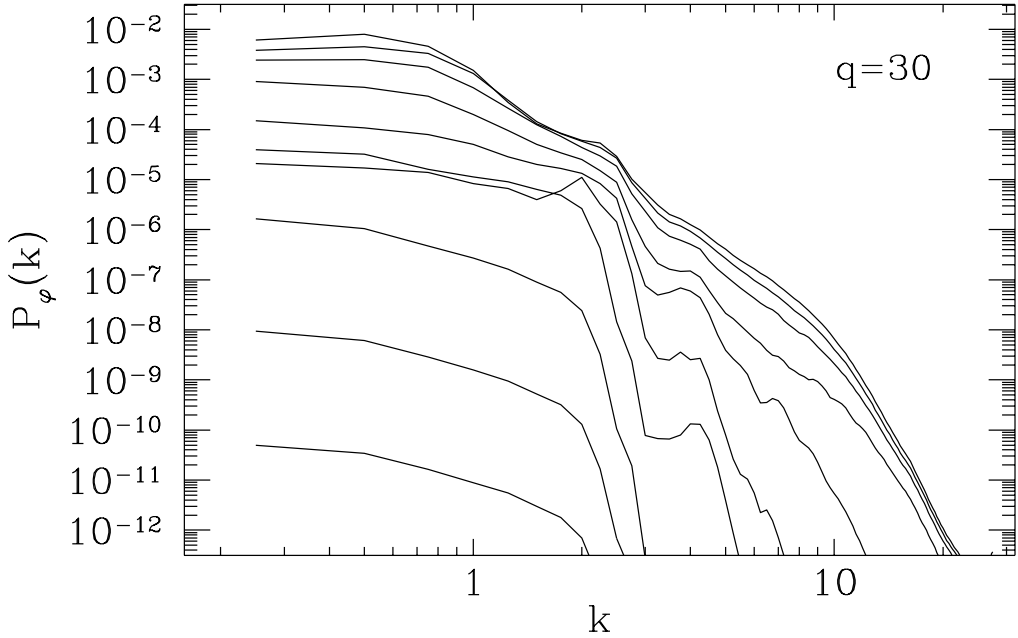,height=3.6in,width=5.6in}
\caption{Power spectrum of the field $\phi$ for the same model as in 
Fig. 1, output every period at the maxima of $\varphi_0(\tau)$; $k$ 
is rescaled comoving momentum (see text).}.
\label{fig:Fig2}
\end{figure}

\begin{figure}
\psfig{file=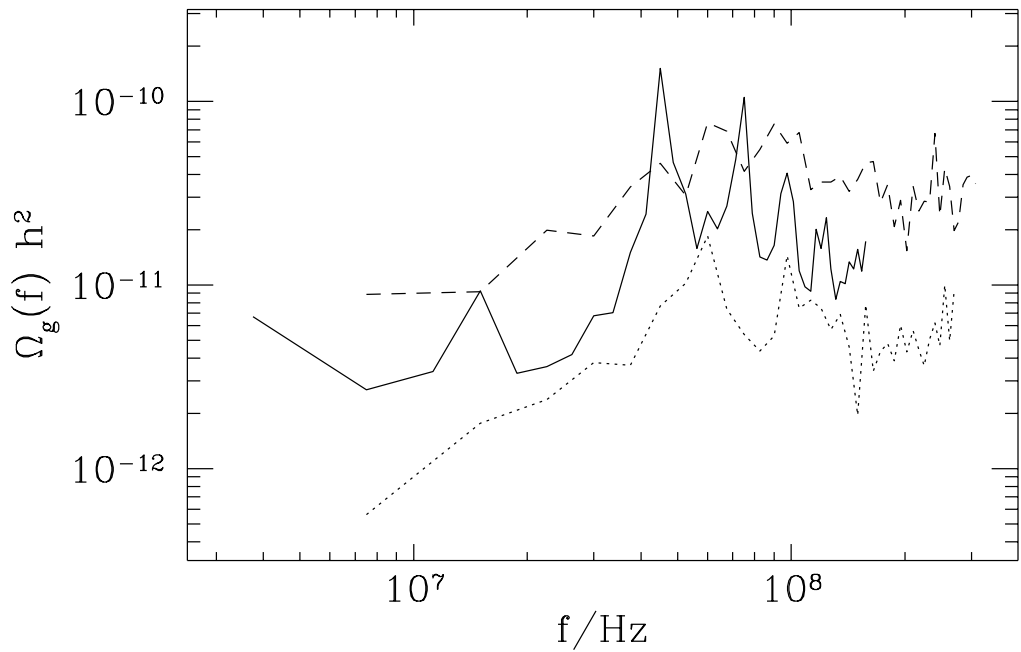,height=3.6in,width=5.6in}
\caption{Today's spectral density of gravitational waves in 
the pure $\lambda \phi^4$ model (solid line) and in the model 
where interaction $g^2 \phi^2 X^2$ with 
a massless scalar field $X$ is added; the dashed line corresponds to 
$q=30$, and the dotted line corresponds to $q=105$, where
$q =g^2/4\lambda $. We used $g_{*}/g_{0}=100$. }
\label{fig:Fig3}
\end{figure}

\end{document}